\documentclass[aps,preprint,prd,showpacs,nofootinbib]{revtex4}

\usepackage{amsmath,amssymb}
\usepackage{graphicx,subfigure}
\usepackage{color,multirow}
\usepackage[colorlinks,linkcolor=magenta,anchorcolor=cyan,citecolor=blue,plainpages=false]{hyperref}

\hypersetup{colorlinks=true,
    breaklinks=true,
    pdfstartview=Fit,
    linkcolor=blue,
    citecolor=blue,
    urlcolor=blue}

\def\lf{\left}
\def\rt{\right}

\def\be{\begin{equation}}
    \def\ee{\end{equation}}
\def\ba{\begin{eqnarray}}
    \def\ea{\end{eqnarray}}

\begin{document}

\title{A fraction of dark matter faded with early dark energy ?}

\author{Hao Wang$^{1,2} $\footnote{\href{wanghao187@mails.ucas.ac.cn}{wanghao187@mails.ucas.ac.cn}}}
\author{Yun-Song Piao$^{1,2,3,4} $ \footnote{\href{yspiao@ucas.ac.cn}{yspiao@ucas.ac.cn}}}

    \affiliation{$^1$ School of Fundamental Physics and Mathematical
        Sciences, Hangzhou Institute for Advanced Study, UCAS, Hangzhou
        310024, China}

    \affiliation{$^2$ School of Physics Sciences, University of
        Chinese Academy of Sciences, Beijing 100049, China}

    \affiliation{$^3$ International Center for Theoretical Physics
        Asia-Pacific, Beijing/Hangzhou, China}

    \affiliation{$^4$ Institute of Theoretical Physics, Chinese
        Academy of Sciences, P.O. Box 2735, Beijing 100190, China}

    \begin{abstract}

In pre-recombination early dark energy (EDE) resolutions of the
Hubble tension, the rise of Hubble constant value $H_0$ is usually
accompanied with the exacerbation of so-called $S_8$ tension.
Inspired by the swampland conjecture, we investigate what if a
fraction $f_*$ of dark matter is coupled to EDE, $m_{cdm}\sim
\exp{\lf(-c{|\Delta\phi_{ede}|\over M_{pl}}\rt)}$ with $c\sim
{\cal O}(1)$. We perform the MCMC analysis for the relevant EDE
models with PlanckCMB, BAO, Pantheon and SH0ES dataset, as well as
DES-Y1 data, and find that such a fraction helps to alleviate the
$S_8$ tension. However, though $c\gtrsim 0.1$ is allowed for a
very small $f_*$, which suggests that a small fraction of dark
matter has ever faded with EDE, $c\sim0$ is also consistent.

    \end{abstract}

    \maketitle
    \tableofcontents

\section{Introduction}
There is a $5\sigma$ conflict between the Hubble constant $H_0\sim
67$km/s/Mpc inferred by Planck collaboration \cite{Planck:2018vyg}
using cosmological microwave background (CMB) data based on
$\Lambda$CDM model and that obtained by SH0ES in light of
Cepheid-calibrated SN data, $H_0\sim 73$km/s/Mpc
\cite{Riess:2021jrx}, which is so-called Hubble tension
\cite{Verde:2019ivm,Riess:2019qba}, see
\cite{Knox:2019rjx,Perivolaropoulos:2021jda,DiValentino:2021izs}
for reviews. Currently, it seems impossible to explain this
conflict by systematic errors, thus it has been widely thought
that this tension signals new physics beyond $\Lambda$CDM.

The Hubble tension is possibly resolved with early dark energy
(EDE) \cite{Karwal:2016vyq,Poulin:2018cxd,Smith:2019ihp}. Here,
the EDE is non-negligible only before recombination, which
suppressed the sound horizon and so naturally brings a high $H_0$
without spoiling fit to CMB and baryon acoustic oscillations (BAO)
data. In particular, if an Anti-de Sitter (AdS) phase existed
around recombination (AdS-EDE model \cite{Ye:2020btb,Ye:2020oix}),
we would have the bestfit $H_0\simeq 73$km/s/Mpc, which is
$1\sigma$ consistent with local $H_0$ measurement. Recently,
besides Planck data, combined analysis of CMB data have been also
performed for EDE, such as Planck+SPT
\cite{Chudaykin:2020acu,Chudaykin:2020igl,Jiang:2021bab},
Planck+ACT DR4 \cite{Hill:2021yec,Poulin:2021bjr}, Planck+ACT
DR4+SPT-3G
\cite{LaPosta:2021pgm,Smith:2022hwi,Jiang:2022uyg,Cruz:2022oqk}
and Planck+BICEP/Keck \cite{Ye:2022afu}.

In such pre-recombination EDE resolutions, the rise of $H_0$ is
usually accompanied with the exacerbation of so-called $S_8$
tension \cite{Hill:2020osr,Ivanov:2020ril,DAmico:2020ods}, see
also \cite{Krishnan:2020obg,Nunes:2021ipq}, where \be
S_8=\sigma_8\lf({\Omega_{m }/ 0.3}\rt)^{1/2},\ee and $\sigma_8$ is
the amplitude of matter perturbations at $8h^{-1}$Mpc scale. It is
well-known that $S_8\sim 0.82$ for $\Lambda$CDM and $S_8\gtrsim
0.84$ for EDE, while the local large-scale structure observations
\cite{Asgari:2019fkq,KiDS:2020suj,DES:2021wwk} have reported lower
$S_8\sim 0.76$. Recently, the resolution of $S_8$ tension has been
intensively studied, which might be completely independent of EDE,
such as Dark matter (DM)-DE drag \cite{Poulin:2022sgp} at low
redshifts, ultra-light axion as a little part of DM
\cite{Allali:2021azp,Ye:2021iwa,Alexander:2022own}, Decaying DM
\cite{FrancoAbellan:2021sxk,Clark:2021hlo,Simon:2022ftd}, massive
neutrino \cite{Reeves:2022aoi}, and also
\cite{Chacko:2016kgg,Buen-Abad:2022kgf}.

However, the DM physics responsible for the $S_8$ tension might
also be relevant with EDE, e.g.
\cite{Karwal:2021vpk,McDonough:2021pdg,Berghaus:2022cwf}. The
conformal coupling of DM to EDE has been considered in
Ref.\cite{Karwal:2021vpk}, see also
\cite{Sakstein:2019fmf,CarrilloGonzalez:2020oac} for
neutrino-assisted EDE. The coupled EDE and the impact of massive
neutrinos also has been studied in
Ref.\cite{Gomez-Valent:2022bku}. The evolution of our Universe
must be implemented in a UV-complete effective field theory (EFT).
It has been argued in Ref.\cite{Ooguri:2006in} that such EFTs must
satisfy the \textit{swampland distance conjecture} (SDC): the
excursion of any field must comply with $|\Delta\phi|\lesssim
M_{pl}$, or it will cause the exponential suppression of the mass
of other fields in EFT. Thus it is possible that DM might be
exponentially lightened (called ``fading dark matter"
\cite{Agrawal:2019dlm}) with the evolution of EDE. However, in
such an early dark sector \cite{McDonough:2021pdg}, the results
favored by current datasets seem conflicted with SDC.

It might be also possible that not all but only a fraction of DM
coupled EDE. In this paper, we will investigate such a coupling in
Axion-like EDE and AdS-EDE models, respectively. In section-II, we
comment the correlation of $S_8-H_0$, and outline our setup in
section-III. In sections-IV and V, we perform the Markov Chain
Monte Carlo (MCMC) analysis with PlanckCMB, BAO, Pantheon and
SH0ES dataset, as well as full DES-Y1 dataset, and report our
results. We conclude in sections-VI.

\section{$S_8-H_0$ in EDE }

It is well-known that Planck dataset strictly set the angular
scale \be \theta_{CMB}={r_s\over D_A}\sim r_s H_0.
\label{theta}\ee where $D_A=\int_0^{z_*} {dz\over H(z)}$ is the
angular radius to last scattering surface,
$r_s=\int_{z_*}^{\infty}\frac{c_s}{H(z)}dz$ is the sound horizon,
and $z_*$ is the redshift of last scattering. In pre-recombination
EDE setup, $r_s$ is suppressed (see \cite{Smith:2022iax} for
recent result) so that we have a high $H_0$ in light of
(\ref{theta}). Recently, the relevant EDE models have widely
studied
e.g.\cite{Smith:2019ihp,Ye:2020btb,Ye:2020oix,Agrawal:2019lmo,Alexander:2019rsc,Berghaus:2019cls,Lin:2019qug,Kaloper:2019lpl,Lin:2020jcb,Braglia:2020bym,Seto:2021xua,Nojiri:2021dze,Nojiri:2022ski,Sabla:2022xzj,MohseniSadjadi:2022pfz},
and
\cite{Lin:2018nxe,SolaPeracaula:2019zsl,Zumalacarregui:2020cjh,Ballesteros:2020sik,SolaPeracaula:2020vpg,Braglia:2020auw},
see also its effects on cosmic birefringence
\cite{Fujita:2020ecn,Murai:2022zur} and gravitational waves
background \cite{Braglia:2021fxn,Chang:2021yog}.

It has been showed in Ref.\cite{Wang:2022jpo} that even if the
state equation $w(z)$ of DE after recombination evolved with the
redshift $z$, its rise for the bestfit ${H_0}$ is also negligible.
However, though the post-recombination beyond-$\Lambda$CDM
modification seems difficult to resolve the Hubble tension, it is
still worth exploring,
e.g.\cite{DiValentino:2017zyq,Vagnozzi:2019ezj,DiValentino:2019ffd,Yang:2021flj,Yang:2020ope,Liu:2021mkv,Alestas:2021luu,Krishnan:2021dyb,Perivolaropoulos:2021bds,Heisenberg:2022gqk,Nunes:2022bhn,Dainotti:2021pqg,Dainotti:2022bzg,Schiavone:2022shz},
see also
\cite{Odintsov:2022eqm,Odintsov:2022umu,Oikonomou:2022tjm} (the
physics of our Universe might be abruptly interrupted at redshifts
$z=0.01$ in the past), or it might be also possible that flat
$\Lambda$CDM model is breaking down
\cite{Krishnan:2020vaf,Colgain:2022nlb,Colgain:2022rxy}.

\begin{figure}[htbp]
{\includegraphics[width=0.7\textwidth]{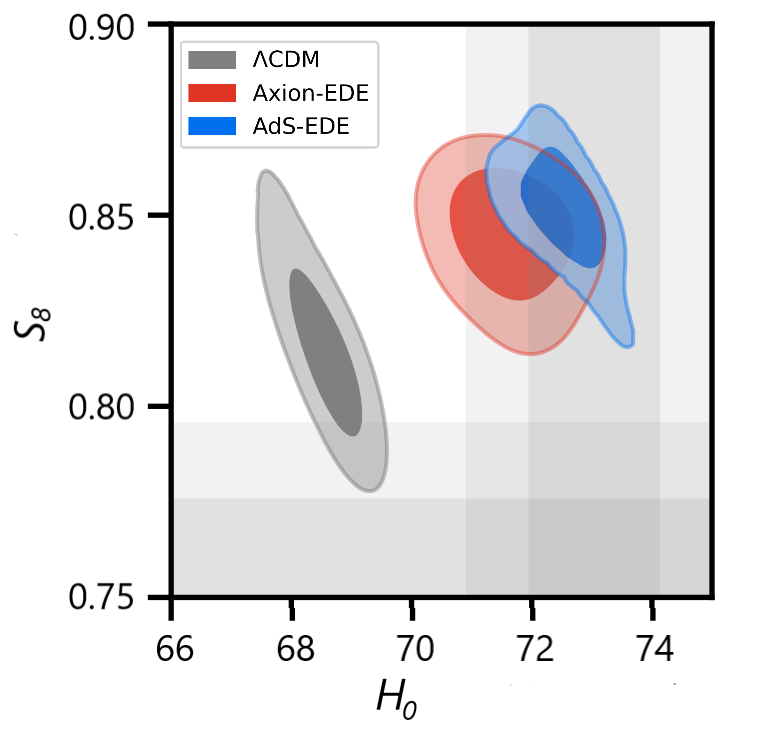}}
\caption{\label{fig0} The $S_8-H_0$ contour for $\Lambda$CDM,
Axion-like EDE and AdS-EDE models. The shadows correspond to the
1$\sigma$ and 2$\sigma$ regions of $H_0$ in light of recent SH0ES
$H_0=73.04\pm1.04$km/s/Mpc \cite{Riess:2021jrx} and $S_8$ in light
of KiDS+VIKING-450+DES-Y1 constraint $S_8=0.755\pm0.02$
\cite{Asgari:2019fkq}, respectively. It is clearly seen that the
EDE also proportionally lift $S_8$ while lift $H_0$.}
\end{figure}

In pre-recombination EDE resolutions, the cosmological parameters
must shift with $\delta H_0$, particularly the shift of
$\omega_m=\Omega_mh^2$ scales approximately \cite{Ye:2020oix} \be
\delta \omega_{m}\simeq 2\frac{\delta
H_0}{H_0}\omega_{m},\label{H0omegam}\ee since PlanckCMB+BAO
dataset required $\omega_m \sim H_0^{2}$ (or $\Omega_m\sim
const.$). The dust-like matter will cluster in the
matter-dominated era, so higher $H_0$ will proportionally bring a
higher $S_8$. Thus the EDE will inevitably suffer from the
exacerbation of $S_8$ tension, see Fig.\ref{fig0}.

In the standard EDE+$\Lambda$CDM model, all DM not only
participated in the background evolution of the Universe but also
is responsible for perturbation growth, which naturally results in
(\ref{H0omegam}) and so suggests exacerbated $S_8$ tension for
EDE. There might, however, other possible matter forms or coupling
which can break the correlation between $S_8$ and $\omega_m$. In
this sense, the $S_8$ tension is actually an opportunity of
understanding CDM physics.

\section{Dark matter fractionally coupled to EDE}

The SDC suggests \cite{Ooguri:2006in} that any EFT is only valid
in field space bounded by the Planck scale, $|\Delta\phi|<
M_{pl}$, and its breakdown that occurs at Planckian field
excursions is encoded in the mass spectrum of other fields, $m\sim
\exp{(-c|\Delta\phi|/M_{pl})}$, where $c\simeq {\cal O}(1)$, see
\cite{Brennan:2017rbf,Palti:2019pca} for reviews.

\begin{figure}[htbp]
        {\includegraphics[width=0.8\textwidth]{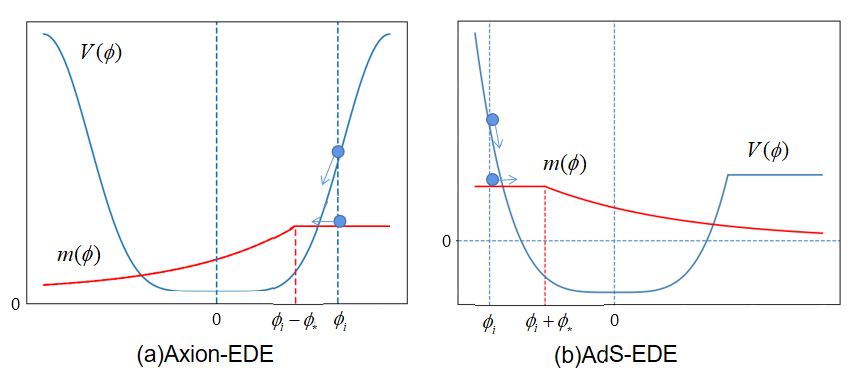}}
\caption{\label{fig1}A sketch of the EDE potential $V(\phi)$ and
$m(\phi)$ in Axion-like EDE and AdS-EDE models, respectively.
Initially the field sits at $\phi_i$, after its excursion
$|\Delta\phi|>\phi_*$, the mass of DM will be exponentially
lightened with the evolution of $\phi$. }
    \end{figure}

Inspired by the SDC, we consider the couple of DM with EDE. The
dark matter is modelled as a population of non-relativistic Dirac
fermions $\psi$, \be {\cal L}_{int}\sim
 -m_{cdm}(\phi)\bar{\psi} \psi,\label{L}\ee \be
 m_{cdm}(\phi)=f_*m(\phi)+(1-f_*)m_i,\quad with \quad 0\leqslant f_*\leqslant 1, \label{mdm}\ee
 \be
m(\phi)=m_i e^{-c({|\Delta\phi|-\phi_*\over M_{pl}})},\quad for
\quad |\Delta\phi|\geq\phi_*.
    \label{mphi}\ee
where $m_i=const$ is the initial mass of DM,
$|\Delta{\phi}|=|\phi-\phi_i|$ (see Fig.\ref{fig1}), $\phi_*$
signals the insensitivity of DM on a shift of $\phi$ within
$|\Delta{\phi}|<\phi_*$, and $c$ is the coupling intensity. Here,
when $c=0$, we have $m_{cdm}=m_i$ (the standard EDE+$\Lambda$CDM
model is recovered).

In non-relativistic limit, we have $\rho_{cdm}= nm_{cdm}(\phi)$,
so \be \rho_{cdm}= nf_*m(\phi)+n(1-f_*)m_i,\ee with $n$ being the
number density, which suggests that $f_*$ is actually equivalent
to the fraction of DM coupled to EDE.

Here, we follow Ref.\cite{McDonough:2021pdg}. The evolution of EDE
is rewitten as $\phi^{\prime\prime}+2{\cal
H}\phi^\prime+a^2V_{\phi}=-a^2\frac{d\rho_{cdm}}{d\phi}$, while
the continuity equation for DM is
    \be
    \rho_{cdm}^\prime+3{\cal H}\rho_{cdm}=\phi^\prime\frac{d\rho_{cdm}}{d\phi},
    \label{rhocdm}\ee
where the \textit{prim}e is the derivative with respect to
$d\eta=dt/a$, and ${\cal H}={a^\prime/a}$. Integrating
Eq.(\ref{rhocdm}), we have
    \ba
\rho_{cdm}(a)=
\frac{3M_{pl}^2H_0^2\Omega_{cdm}}{a^3}[1-f(\phi_0)+\frac{m(\phi)}{m(\phi_0)}f(\phi_0)],
\label{rhocdm1}\ea where
$f(\phi)=\frac{{m(\phi)}f_*}{{m(\phi)}f_*+m_i(1-f_*)}$, and
$\phi_0$ is the present-day value of $\phi$. In axion-like EDE
model, see Fig.\ref{fig1}(a), $\phi-\phi_i<0$ for $\phi_i>0$, so
$d\rho_{cdm}/d\phi=c\rho_{cdm}f(\phi)/{M_{pl}}$. The scalar field
will eventually settle at the bottom of its potential, so
$\phi_0=0$.

In the synchronous gauge, with $\rho_{cdm}$ in Eq.(\ref{rhocdm1}),
the perturbations equations have been derived fully in
Ref.\cite{McDonough:2021pdg}. However, in AdS-EDE model
$\phi-\phi_i>0$, see Fig.\ref{fig1}(b), so
$d\rho_{cdm}/d\phi=-c\rho_{cdm}f(\phi)/{M_{pl}}$. This suggests
that $\phi_0$ must be obtained by solving the equation of motion,
so it is not convenient to use Eq.(\ref{rhocdm1}). Integrating
Eq.(\ref{rhocdm}), instead we have \ba \rho_{cdm}^{(AdS)}(a)&=&
\frac{3M_{pl}^2H_0^2{\Omega}_{cdm}}{a^3}[1-f(\phi_0)+(1-f(\phi_0)){f_*\over
1-f_*} \frac{m(\phi)}{m(\phi_{i})}]\nonumber\\
&=&
\frac{3M_{pl}^2H_0^2{\tilde\Omega}_{cdm}}{a^3}[1+f_*^{(AdS)}\frac{m(\phi)}{m(\phi_{i})}],
\label{rhocdm3}\ea where ${\tilde
\Omega}_{cdm}=\Omega_{cdm}(1-f(\phi_{0}))$ is defined to absorb
$\phi_0$ and $f_*^{(AdS)}= f_*/(1-f_*)$.

\section{Dataset and results}

Here, our baseline dataset consists of:

1. \textbf{CMB}: Planck 2018 low-l and high-l TT, TE, EE spectra ,
and reconstructed CMB lensing spectrum
\cite{Planck:2018vyg,Planck:2018lbu,Planck:2019nip}.

2. \textbf{BAO}: The BOSS DR12\cite{BOSS:2016wmc} with its full
covariant matrix for BAO as well as the 6dFGS
\cite{Beutler:2011hx} and MGS of SDSS \cite{Ross:2014qpa}.

3. \textbf{Supernovae}: The Pantheon dataset
\cite{Pan-STARRS1:2017jku}.

4. \textbf{SH0ES}: To avoid the \textit{prior volume effect}
\footnote{In AdS-EDE model, the prior volume effect is actually
removed by AdS bound, as explained in
\cite{Jiang:2021bab,Ye:2022afu}.}
\cite{Schoneberg:2021qvd,Herold:2021ksg,Gomez-Valent:2022hkb},
which will compel the EDE models prefer a low $f_{ede}$, we take
$H_0=73.04\pm1.04$km/s/Mpc reported by the SH0ES
\cite{Riess:2021jrx} as the Gaussian prior, see also
\cite{Camarena:2021jlr,Efstathiou:2021ocp}.

We modified the MontePython-3.3 sampler
\cite{Audren:2012wb,Brinckmann:2018cvx} and CLASS codes
\cite{Lesgourgues:2011re,Blas:2011rf}\footnote{The corresponding
cosmological codes are available at: axion-like EDE
(\url{https://github.com/PoulinV/AxiCLASS}) and AdS-EDE
(\url{https://github.com/genye00/class_multiscf}).} to perform the
MCMC analysis for axion-like EDE and AdS-EDE, respectively, with
baseline dataset and baseline+DES-Y1 dataset, see
\cite{DES:2017myr} for DES-Y1 data. The Gelman-Rubin criterion for
all chains is converged to $R-1<0.05$.

\subsection{Axion-like EDE}

\begin{figure}[htbp]
    {\includegraphics[width=0.8\textwidth]{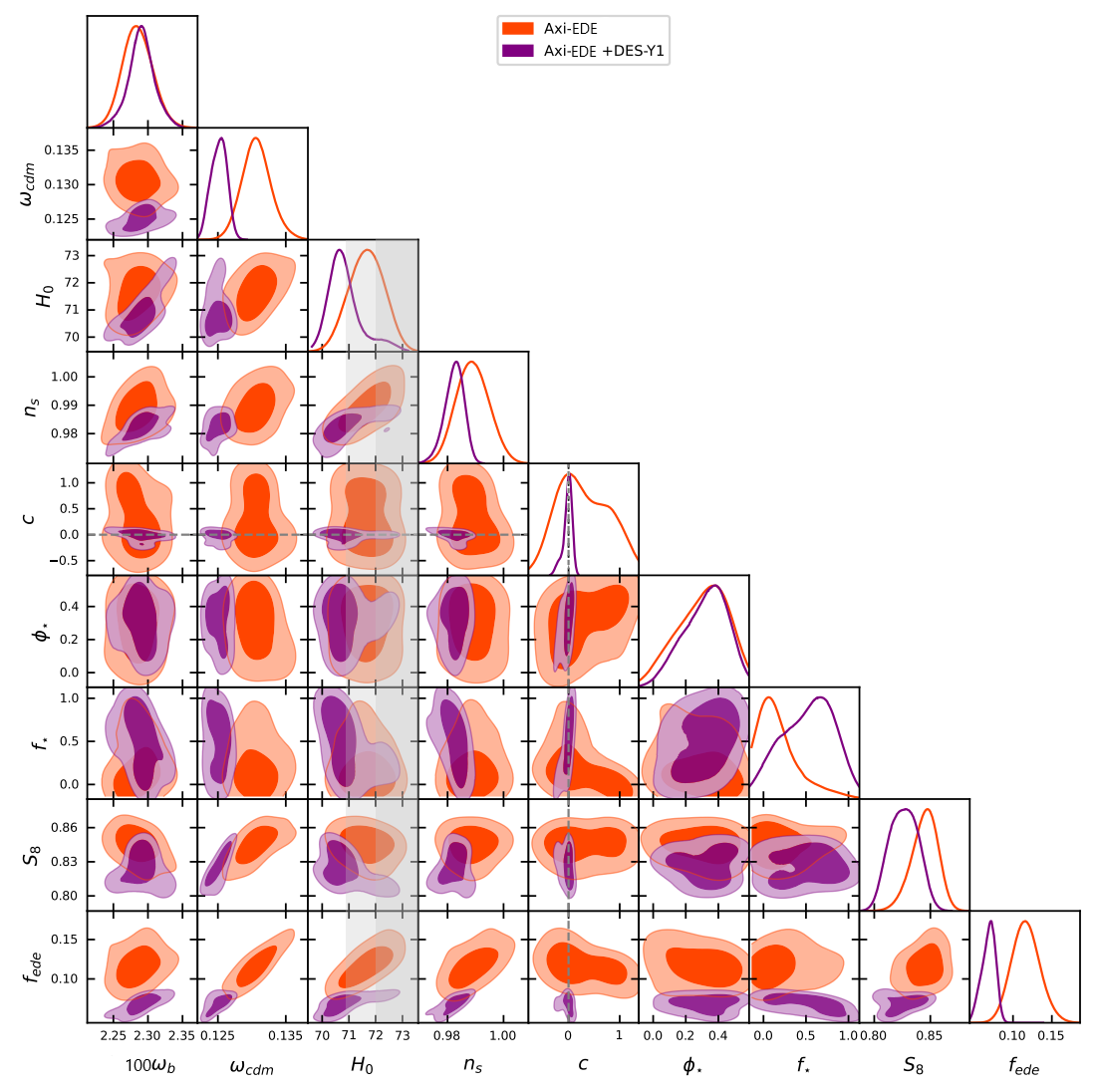}}
\caption{\label{fig2} Posterior distributions for Axion-like EDE
with coupling (\ref{mphi}) in fit to the baseline and the baseline
dataset+DES-Y1 datasets, respectively. The shadows correspond to
the 1$\sigma$ and 2$\sigma$ regions of $H_0$ in light of recent
SH0ES \cite{Riess:2021jrx}.}
    \end{figure}

The original EDE model is: axion-like EDE \cite{Poulin:2018cxd}.
An axion field with $V(\phi)\propto(1-cos[\phi/f])^3$ is
responsible for EDE (see recent
\cite{McDonough:2022pku,Kojima:2022fgo} for models in string
theory), which starts to oscillate at the redshift $z_c\sim 3000$.
It is usually parameterized by $\phi_i$, $a_c$ and $f_{ede}$
\cite{Poulin:2018cxd,Smith:2019ihp}. In Table.\ref{tab1}, we
present the MCMC results for axion-like EDE  with the baseline
dataset and baseline+DES-Y1 dataset. In Fig.\ref{fig2}, we show
the 1$\sigma$ and 2$\sigma$ marginalized posterior distributions
of parameter set
$\{\omega_b,\omega_{cdm},H_0,\ln({10^{10}}A_s),n_s,\tau_{reio},{\log_{10}}{a_c},f_{ede},\phi_i,c,\phi_*,f_*\}$.

Though with the baseline dataset, we have the bestfit
$S_8=0.8438$, which is larger than local $S_8$ measurements, the
baseline+DES-Y1 dataset prefers a lower $S_8$ (the bestfit is
$S_8=0.8186$, which almost equals to $S_8=0.8156$ in
$\Lambda$CDM), than that with only baseline dataset. However, the
cost is that the bestfit $H_0=70.14$ is lowered.

In Table.\ref{tab1}, we see that with baseline dataset, $c\sim 0$
at 1$\sigma$ region, consistent with the result in
Ref.\cite{McDonough:2021pdg}, which suggests that such a coupling
(\ref{L}) is not favored, while the case is not altered with
baseline+DES-Y1 dataset. The bestfit of $c$ is negative and
inconsistent with SDC, but the possibility of $c>0$ is not ruled
out.

    \begin{table}[htbp]
        \scalebox{0.85}{
        \begin{tabular}{|c|c|c|c|}
            \hline
            \multirow{2}{*}{Parameters}&$\Lambda$CDM&\multicolumn{2}{|c|}{Axion-EDE}\\\cline{2-4}
            &\multicolumn{2}{c|}{baseline}&{baseline+DES-Y1}\\
            \hline
            100$\omega_b$&$2.252(2.249)\pm0.013$&$2.284(2.286)\pm0.020$&$2.290(2.282)\pm0.017$\\
            $\omega_{cdm}$&$0.1182(0.1184)\pm0.0008$&$0.1306(0.1290)\pm0.0020$&$0.1251(0.1236)\pm0.0011$\\
            $H_0$&$68.21(68.16)\pm0.39$&$71.66(70.87)\pm0.63$&$70.83(70.14)\pm0.57$\\
            ln($10^{10}$$A_s$)&$3.052(3.052)\pm0.015$&$3.060(3.051)\pm0.013$&$3.051(3.031)\pm0.013$\\
            $n_s$&$0.9691(0.9686)\pm0.0035$&$0.9889(0.9834)\pm0.0056$&$0.9824(0.9786)\pm0.0033$\\
            $\tau_{reio}$&$0.0595(0.0594)\pm0.0073$&$0.0576(0.0571)\pm0.0064$&$0.0553(0.0487)\pm0.0047$\\
            $f_{ede}$&-&$0.116(0.101)\pm0.017$&$0.068(0.055)\pm0.008$\\
            $log_{10}a_c$&-&$-3.748(-3.841)\pm0.137$&$-3.585(-3.602)\pm0.073$\\
            \hline
            $c$&-&$0.289(-0.129)\pm0.472$&$-0.015(-0.012)\pm0.082$\\
            $\phi_*$&-&$0.305(0.361)\pm0.147$&$0.321(0.406)\pm0.127$\\
            $f_*$&-&$0.183(0.222)\pm0.229$&$0.521(0.860)\pm0.289$\\
            \hline
            $S_8$&$0.8140(0.8156)\pm0.0098$&$0.8451(0.8438)\pm0.0112$&$0.8263(0.8186)\pm0.0122$\\
            \hline
        \end{tabular}}
\caption{\label{tab1}Mean(best-fit) values of $\Lambda$CDM and
Axion-like EDE with coupling (\ref{mphi}) in fit to the baseline
and the baseline+DES-Y1 datasets, respectively.}
    \end{table}

\subsection{AdS-EDE}

In AdS-EDE model \cite{Ye:2020btb}, we have
$V(\phi)=V_0(\phi/M_p)^{4}-V_{ads}$, see Fig.\ref{fig1}(b), which
is glued to a cosmological constant $V(\phi)=\Lambda$ by
interpolation ($V_{ads}>0$ is the AdS depth). Here, the scalar
field starts to roll at the redshift $z_c\sim 3000$, and then
rolls over an AdS minimum like a fluid with $w>1$. It climbs up to
the $\Lambda> 0$ region shortly after recombination, hereafter the
Universe will be effectively described by the $\Lambda$CDM model.
It is well-known that AdS vacua are ubiquitous in string
landscape, so the AdS-EDE model can be well-motivated, see also
\cite{Akarsu:2019hmw,Visinelli:2019qqu,Dutta:2018vmq,Calderon:2020hoc,Akarsu:2021fol,Sen:2021wld,DiGennaro:2022ykp,Moshafi:2022mva}
for other studies on the implications of AdS vacua for our
Universe.

\begin{figure}[htbp]
    {\includegraphics[width=0.8\textwidth]{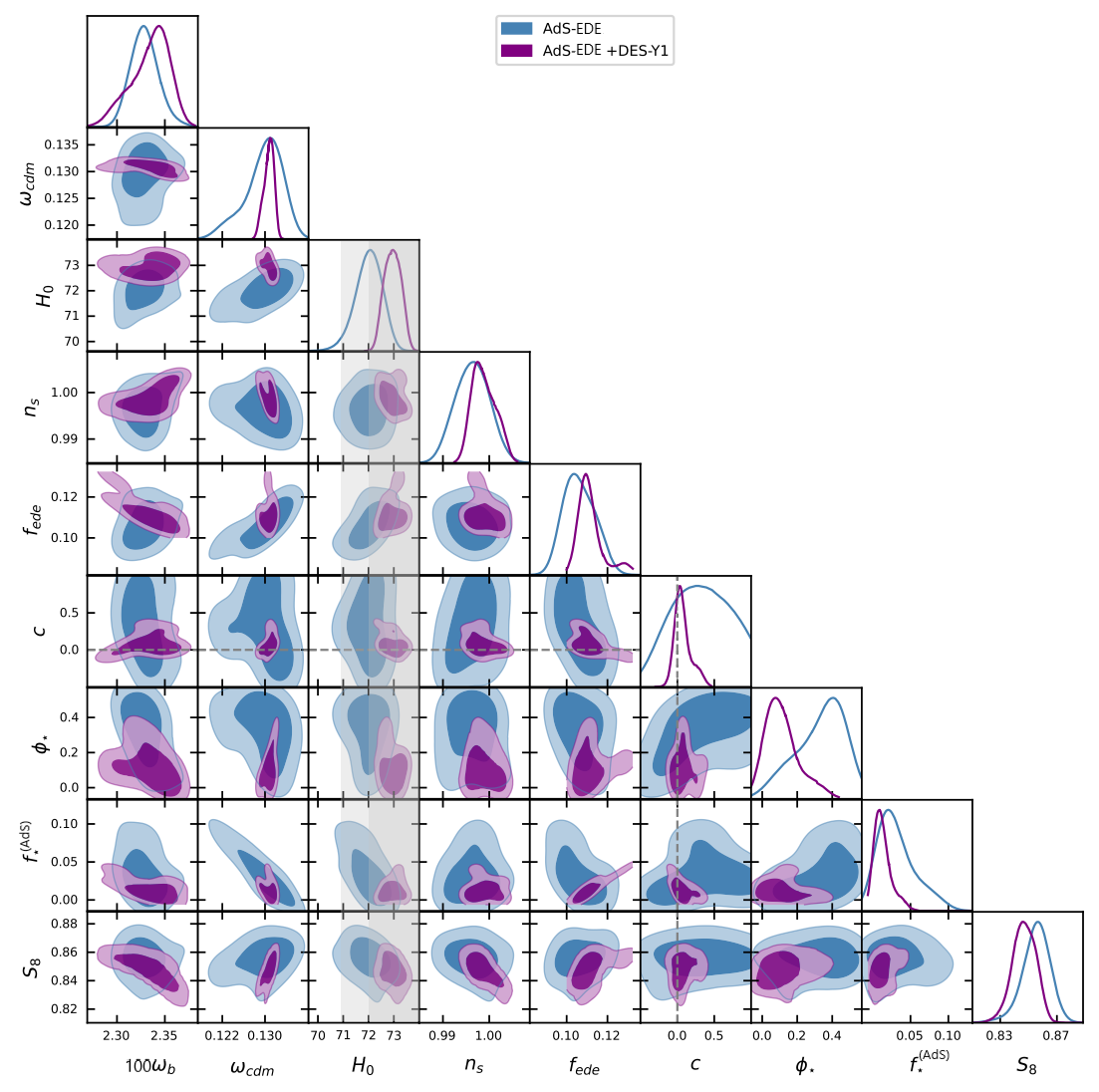}}
\caption{\label{fig3}Posterior distributions for AdS-EDE with
coupling (\ref{mphi}) in fit to the baseline and the
baseline+DES-Y1 datasets, respectively. The shadows correspond to
the 1$\sigma$ and 2$\sigma$ regions of $H_0$ in light of recent
SH0ES \cite{Riess:2021jrx}.}
    \end{figure}

The AdS-EDE model is usually parameterized by $V_{ads}$, $z_c$ and
$f_{ede}$. In order to have a significant AdS phase while make the
field able to climb out of the AdS well, we fixed $V_{ads}$ by
setting $ V_{ads}=0.26\times 10^{4}
\left(\rho_\text{m}(z_c)+\rho_\text{r}(z_c)\right)$, as in
Ref.\cite{Ye:2020btb}. In Table.\ref{tab2}, we present the MCMC
results for AdS-EDE with the baseline dataset and baseline+DES-Y1
dataset. In Fig.\ref{fig3}, we show the 1$\sigma$ and 2$\sigma$
marginalized posterior distributions of parameter set
$\{\omega_b,\omega_{cdm},H_0,\ln({10^{10}}A_s),n_s,\tau_{reio},\ln(1+z_c),
f_{ede},c,\phi_*,f_*\}$.

Though the baseline+DES-Y1 dataset prefers a lower $S_8$ (the
bestfit is $S_8=0.8433$) than that with only baseline dataset, it
is still larger than that in $\Lambda$CDM. However, unlike in
axion-like EDE, $f_{ede}$ is not suppressed by the coupling
(\ref{L}), since $f_*\sim 0.02$ is fairly small. It is also noted
that with baseline+DES-Y1 dataset, we have the bestfit
$H_0=73.33$, which is slightly higher than that in AdS-EDE
\cite{Ye:2020btb}.

In Table.\ref{tab1}, we see that with baseline dataset, $c\sim
0.4$ at 1$\sigma$ region, which is different from that in
axion-like EDE (see section-III.A) and consistent with SDC, and
with baseline+DES-Y1 dataset $c\gtrsim 0.1$ at 1$\sigma$ region.
However, in both case $c\sim 0$ is still 1$\sigma$ consistent. In
AdS-EDE, $f_*\sim 0.03$ is smaller than that in axion-like EDE,
and $f_*=1$ is ruled out at $2\sigma$.

    \begin{table}[htbp]
        \scalebox{0.85}{
        \begin{tabular}{|c|c|c|c|}
            \hline
            \multirow{2}{*}{Parameters}&$\Lambda$CDM&\multicolumn{2}{|c|}{AdS-EDE}\\\cline{2-4}
            &\multicolumn{2}{c|}{baseline}&{baseline+DES-Y1}\\
            \hline
            100$\omega_b$&$2.252(2.249)\pm0.013$&$2.328(2.327)\pm0.014$&$2.334(2.339)\pm0.019$\\
            $\omega_{cdm}$&$0.1182(0.1184)\pm0.0008$&$0.1298(0.1299)\pm0.0035$&$0.1307(0.1302)\pm0.0009$\\
            $H_0$&$68.21(68.16)\pm0.39$&$72.01(72.05)\pm0.51$&$72.95(73.33)\pm0.32$\\
            ln($10^{10}$$A_s$)&$3.052(3.052)\pm0.015$&$3.076(3.088)\pm0.014$&$3.086(3.095)\pm0.013$\\
            $n_s$&$0.9691(0.9686)\pm0.0035$&$0.9963(0.9973)\pm0.0035$&$0.9988(0.9967)\pm0.0024$\\
            $\tau_{reio}$&$0.0595(0.0594)\pm0.0073$&$0.0579(0.0601)\pm0.0075$&$0.0624(0.0681)\pm0.0072$\\
            $f_{ede}$&-&$0.1061(0.1002)\pm0.0076$&$0.1114(0.1081)\pm0.0059$\\
            ln$(1+z_c)$&-&$8.2697(8.2138)\pm0.0958$&$8.1570(8.2925)\pm0.0671$\\
            \hline
            $c$&-&$0.367(0.302)\pm0.434$&$0.069(0.035)\pm0.109$\\
            $\phi_*$&-&$0.333(0.323)\pm0.136$&$0.113(0.023)\pm0.092$\\
            $f_*$&-&$0.030(0.008)\pm0.025$&$0.013(0.015)\pm0.0180$\\
            \hline
            $S_8$&$0.8140(0.8156)\pm0.0098$&$0.8554(0.8610)\pm0.0097$&$0.8470(0.8433)\pm0.0086$\\
            \hline
        \end{tabular}}
\caption{\label{tab2}Mean(best-fit) values of $\Lambda$CDM and
AdS-EDE with coupling (\ref{mphi}) in fit to the baseline and the
baseline+DES-Y1 datasets, respectively.}
    \end{table}

We list the $\chi^2$ of bestfit points for axion-like EDE and
AdS-EDE models in Tables.\ref{tab3} and \ref{tab4}, respectively.
In Tables.\ref{tab3}, only with baseline dataset, we see that both
models have improvements over the bestfit $\Lambda$CDM by
$\Delta\chi^2\sim{-19}$, where the $\chi^2$ of Planck low-l TT, EE
and $H_0$ are significantly improved while the $\chi^2$ of BAO is
slightly exacerbated. In Table.\ref{tab4} with baseline+DES-Y1
dataset, we see that both models have improvements over the
bestfit $\Lambda$CDM but by only $\Delta\chi^2\sim{-6}$. In both
axion-like EDE and AdS-EDE models, compared with $\Lambda$CDM, the
$\chi^2$ of DES-Y1 is exacerbated with $\Delta\chi^2\sim 3$ and
$\Delta\chi^2\sim 8$, respectively.

        \begin{table}[htbp]
        \begin{tabular}{|c|c|c|c|c|c|}
            \hline
\multirow{2}{*}{Dataset}&\multirow{2}{*}{$\Lambda$CDM}&\multicolumn{2}{|c|}{Axion-EDE}&\multicolumn{2}{|c|}{AdS-EDE}\\\cline{3-6}
            &&uncoupled & &uncoupled &\\
            \hline
            Planck high-l TT,TE,EE&2347.50&2344.27&2349.29&2347.57&2346.34\\
            Planck low-l EE&398.2&398.19&396.06&397.50&395.78\\
            Planck low-l TT&23.9&20.56&20.92&20.83&20.72\\
            Planck lensing&9.10&10.12&9.46&10.60&9.72\\
            BAO BOSS DR12&1.8&3.46&3.42&2.1&3.53\\
            BAO smallz 2014&2.2&2.06&1.92&2.2&1.75\\
            Pantheon&1026.9&1026.68&1026.87&1026.9&1026.88\\
            SH0ES&15.40&1.38&3.08&0.8&1.05\\
            \hline
            Total&3825&3811.79&3806.04&3808.5&3805.82\\
            $\Delta\chi^2$&0&-13.21&-18.96&-16.5&-19.18\\
            \hline
        \end{tabular}
\caption{\label{tab3}$\chi^2$ of both Axion-like EDE and AdS-EDE
for the baseline dataset, where ``uncoupled" corresponds to the
models without coupling (\ref{mphi}).}
    \end{table}
    \begin{table}[htbp]
            \begin{tabular}{|c|c|c|c|}
                \hline
                Dataset&$\Lambda$CDM&Axion-EDE&AdS-EDE\\
                \hline
                Planck high-l TT,TE,EE&2354.43&2354.31&2357.54\\
                Planck low-l EE&398.09&395.80&395.79\\
                Planck low-l TT&21.94&20.36&20.19\\
                Planck lensing&9.10&10.26&10.58\\
                BAO BOSS DR12&4.56&3.49&3.99\\
                BAO smallz 2014&3.01&2.14&2.67\\
                Pantheon&1027.17&1026.98&1026.93\\
                SH0ES&13.20&7.72&0.45\\
                DSE-Y1&517.73&520.82&525.28\\
                \hline
                Total&4349.27&4342.94&4343.46\\
                $\Delta\chi^2$&0&-6.33&-5.81\\
                \hline
            \end{tabular}
            \caption{\label{tab4}$\chi^2$ of both Axion-like EDE and AdS-EDE
with coupling (\ref{mphi}) for baseline+DES-Y1 dataset.}
    \end{table}

\begin{figure}[htbp]
    \subfigure{\includegraphics[width=0.35\textwidth]{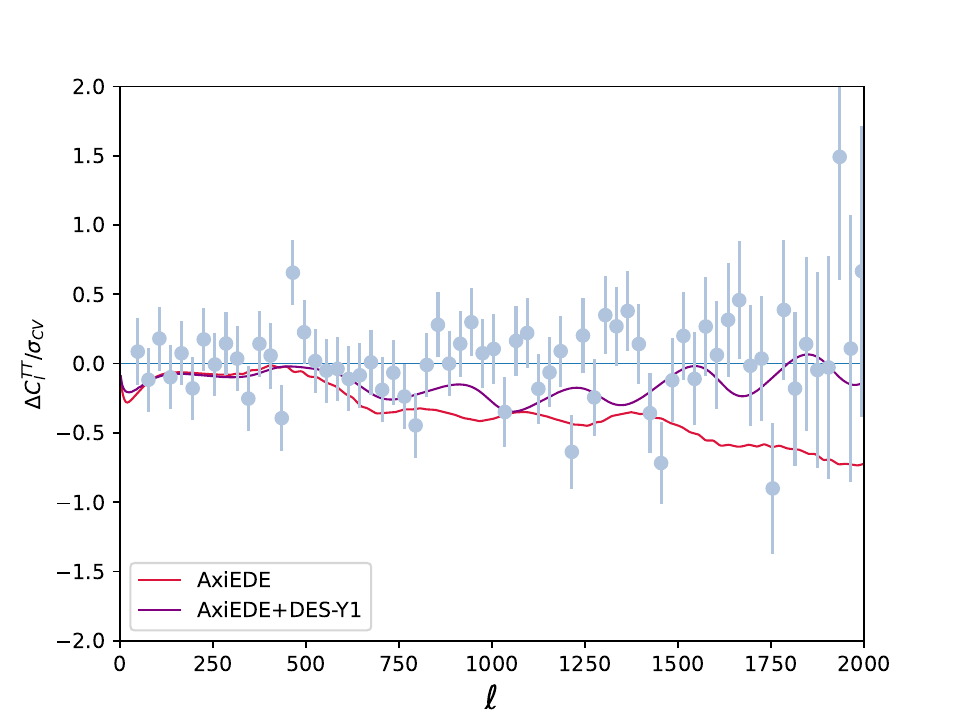}}\label{AxiTT}
    \subfigure{\includegraphics[width=0.35\textwidth]{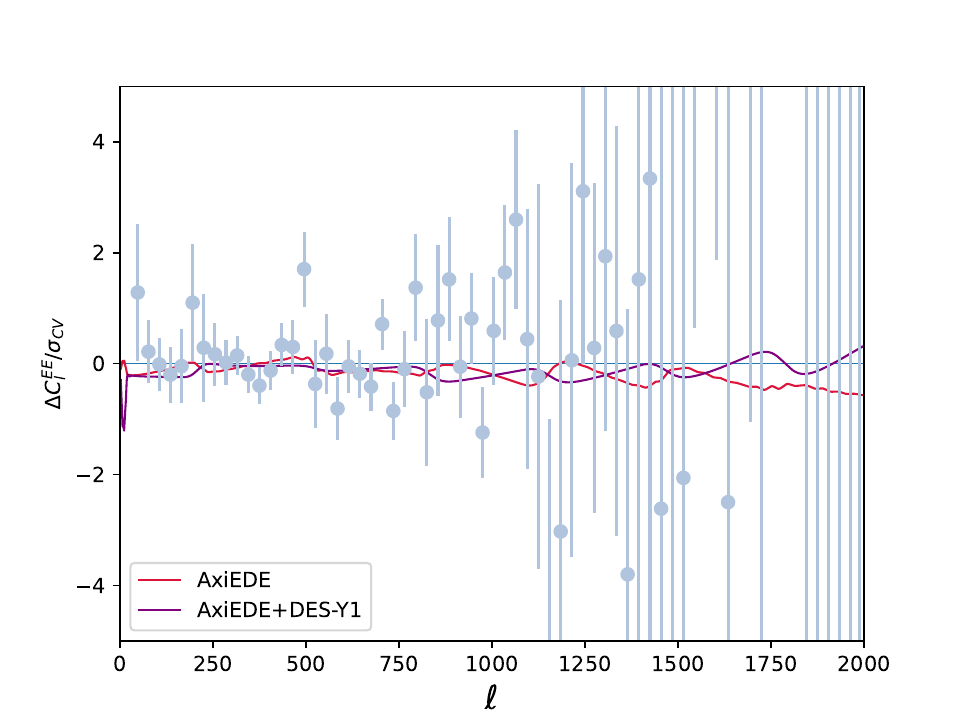}}\label{AxiEE}
    \subfigure{\includegraphics[width=0.35\textwidth]{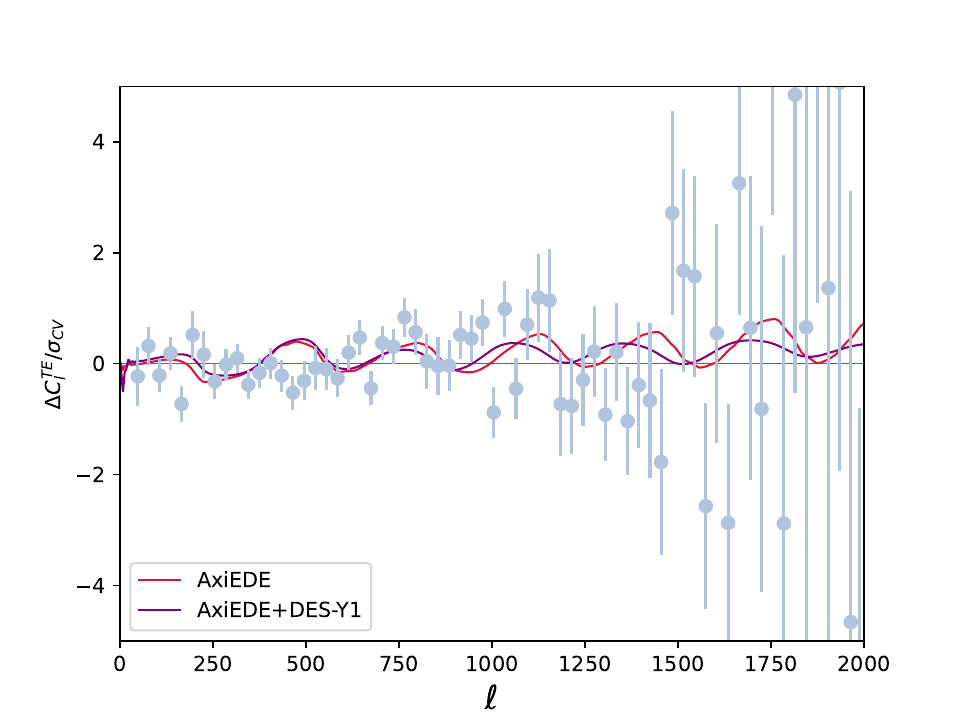}}\label{AxiTE}
\caption{\label{fig5} The TT, EE and TE residuals
${\Delta}C_l/\sigma_{CV}$ for Axion-EDE model with coupling
(\ref{mphi}) in fit to the baseline and baseline+DES-Y1 datasets,
respectively. The reference model is $\Lambda$CDM. }
    \end{figure}

\begin{figure}[htbp]
    \subfigure{\includegraphics[width=0.35\textwidth]{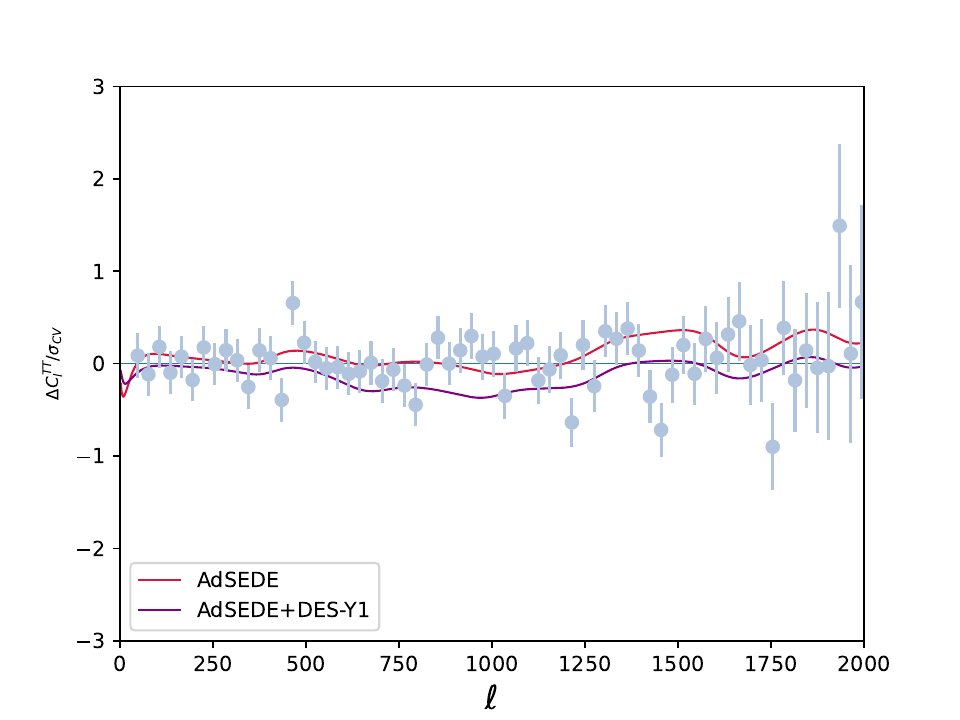}}\label{AdSTT}
    \subfigure{\includegraphics[width=0.35\textwidth]{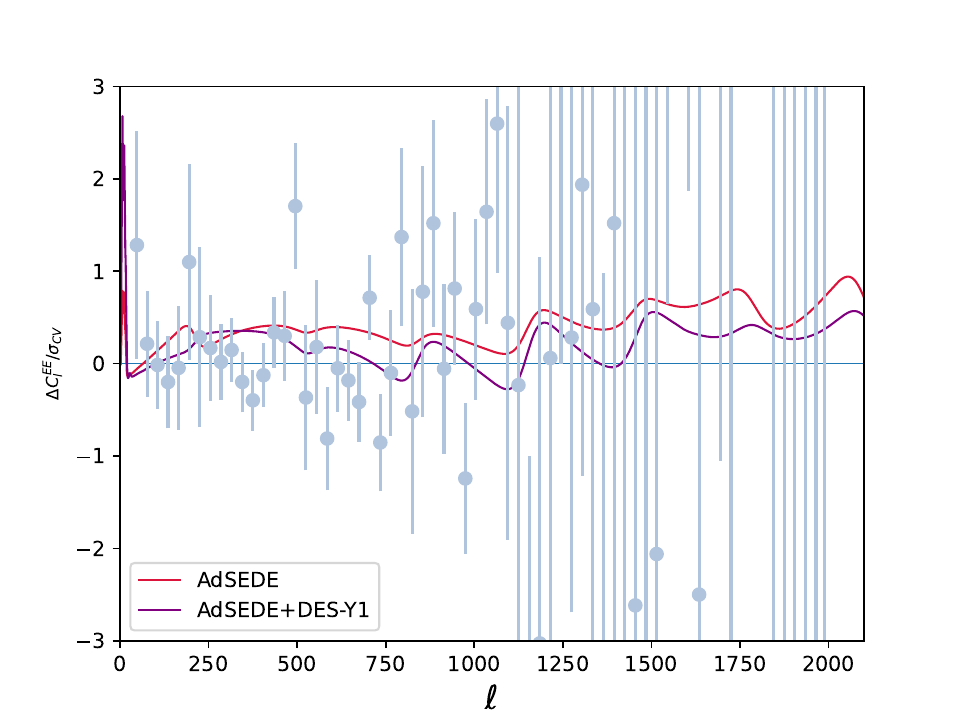}}\label{AdSEE}
    \subfigure{\includegraphics[width=0.35\textwidth]{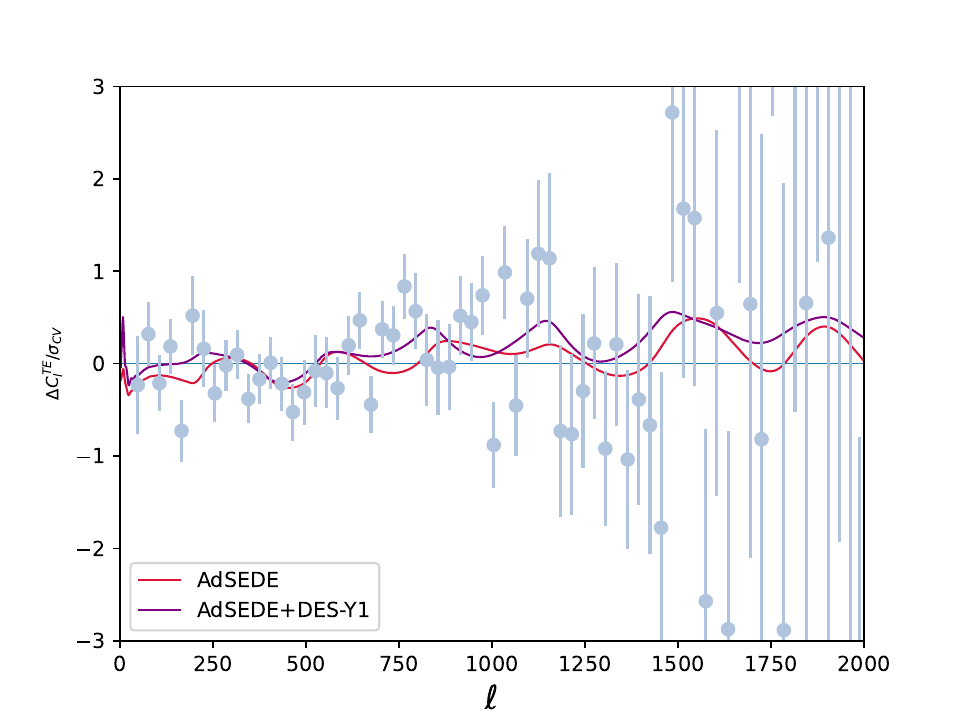}}\label{AdSTE}
\caption{\label{fig6} The TT, EE and TE residuals
${\Delta}C_l/\sigma_{CV}$ for AdS-EDE model with coupling
(\ref{mphi}) in fit to the baseline and baseline+DES-Y1 datasets,
respectively. The reference model is $\Lambda$CDM.}
    \end{figure}

We also plot the TT, EE and TE residuals
${\Delta}C_l=C_{l,model}-C_{l,\Lambda}$ of both models in units of
the cosmic variance per multipole
\[ \sigma_{CV} =\left \{
\begin{array}{rl}
&\sqrt{2/(2l+1)}C^{TT}_l,\quad TT\\
&\sqrt{1/(2l+1)}\sqrt{{C^{TT}_l}{C^{EE}_l}+(C^{TE}_l)^2},\quad TE\\
&\sqrt{2/(2l+1)}C^{EE}_l,\quad EE\\
\end{array}\right. \]
in Figs.\ref{fig5} and \ref{fig6}. The TT residual becomes
comparable to $\sigma_{CV}$ at $l\sim700$ for axion-like EDE (but
is suppressed by DES-Y1 dataset), while DES-Y1 significant impacts
the TT, EE and TE residuals of AdS-EDE.

\section{Has DM ever faded ? }

\begin{figure}[htbp]
\includegraphics[width=0.7\textwidth]{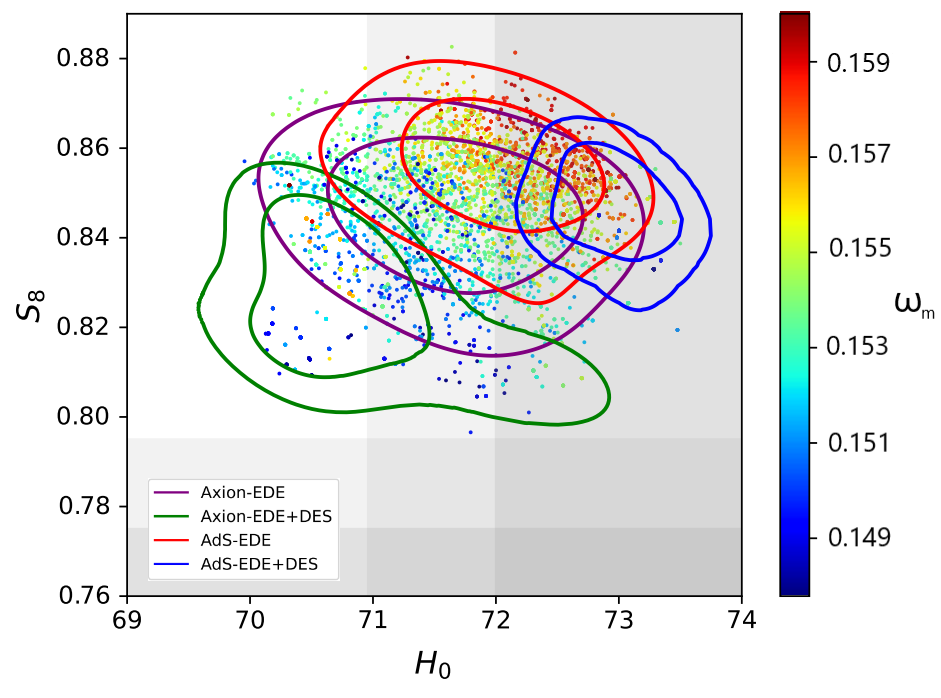}
\caption{\label{fig8}The $S_8-H_0$ contour of Axion-like EDE and
AdS-EDE with the coupling (\ref{mphi}) in fit to the baseline and
baseline+DES-Y1 datasets, respectively. The shadows correspond to
the 1$\sigma$ and 2$\sigma$ regions of $H_0$ in light of recent
SH0ES $H_0=73.04\pm1.04$km/s/Mpc \cite{Riess:2021jrx} and $S_8$ in
light of KiDS+VIKING-450+DES-Y1 constraint $S_8=0.755\pm0.02$
\cite{Asgari:2019fkq}.}
    \end{figure}

In Fig.\ref{fig8}, we see that the smaller $S_8$ caused by DES-Y1
dataset brought with a lower bestfit $\omega_{cdm}=0.1236$ and
$H_0=70.14$ for axion-like EDE, but a higher bestfit
$\omega_{cdm}=0.1302$ and $H_0=73.33$ for AdS-EDE. Thus though
(\ref{H0omegam}) is still right for both models with the coupling
(\ref{mphi}), such a coupling actually impairs the correlation
between $\omega_m$ and $S_8$, so that the rise of $H_0$ must not
be accompanied with the exacerbation of $S_8$ tension.

\begin{figure}[htbp]
        \includegraphics[width=\textwidth]{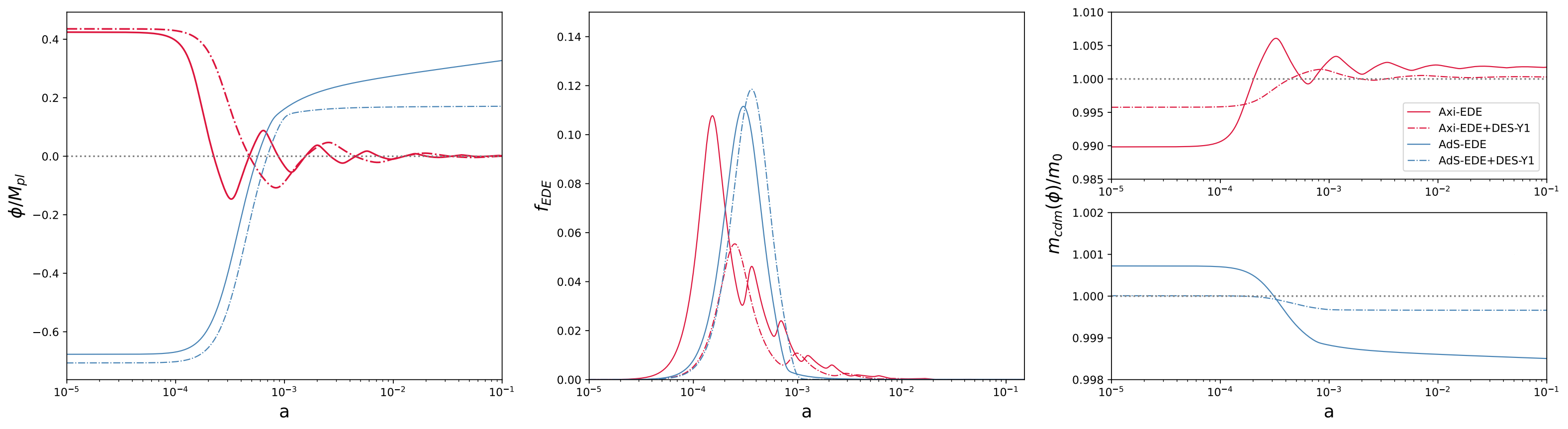}
\caption{\label{fig4} The evolutions of the scalar field,
$f_{ede}$ and $m_{cdm}(\phi)$ in Axion-like EDE and AdS-EDE models
with their bestfit values.}
    \end{figure}

Though in axion-like EDE and AdS-EDE models $c=0$ is 1$\sigma$
consistent, however, the $1\sigma$ contour of $c$ is wide so that
$c\gtrsim 0.1$ is also allowed due to a small $f_*$, see Table.1.
In Fig.\ref{fig4}, we plot the evolution of the scalar field,
$f_{ede}$, and the mass $m_{cdm}(\phi)$ of DM in both models with
their bestfit values. The baseline dataset allows a higher
$c\gtrsim0.1$, so a larger shift of $m_{cdm}(\phi)$, which
suggests that a small fraction of DM ($f_*\sim 0.2$ and $f_*\sim
0.03$ for both models) will fade with EDE. However, it should be
mentioned that in axion-like EDE the bestfit of $c\sim 0$ is a
negative value, consistent with the result in
Ref.\cite{McDonough:2021pdg}.

However, after the DES-Y1 dataset included, we have $c\sim 0$
further, particularly for axion-like EDE with smaller excursion of
scalar field and a lower $f_{ede}$. Thus in both the axion-like
EDE and AdS-EDE models the fading of DM is actually not be favored
by the baseline+DEY-Y1 dataset, but it can not be ruled out at
present.

    \section{Conclusions}

Inspired by SDC, we investigated the impact of DM fractionally
coupled to EDE, specially the possibility of resolving $S_8$
tension. We performed the MCMC analysis for Axion-like EDE and
AdS-EDE, respectively, with PlanckCMB, BAO, Pantheon and SH0ES
dataset (our baseline dataset), as well as DES-Y1 dataset.

It is found that the $S_8$ tension can be alleviated with such a
fractional coupling. In axion-like EDE model, the baseline+DES-Y1
dataset prefers a lower $S_8$ (the bestfit is $S_8\simeq 0.82$,
which almost equals to $S_8=0.8156$ in $\Lambda$CDM), however, the
cost is a lower bestfit $H_0=70.14$, while in AdS-EDE model,
though the baseline+DES-Y1 dataset prefers a lower $S_8$ (the
bestfit is $S_8=0.8433$), it is still larger than that in
$\Lambda$CDM.

The baseline+DES-Y1 dataset allows $c\gtrsim 0.1$ in both EDE
models due to a small $f_*\ll 1$ (bestfit $f\sim 0.5$ for
axion-like EDE while a smaller $f_*\sim 0.02$ for AdS-EDE), which
so is compatible with SDC and suggests that a small fraction of DM
has ever faded with EDE. However, $c\sim 0$ is still at $1\sigma$
region, particularly in axion-like EDE the bestfit of $c$ is a
negative value, consistent with the result in
Ref.\cite{McDonough:2021pdg}.

It is worth mentioning that in EDE models with
fullPlanck+BAO+Pantheon dataset, the shift of primordial scalar
spectral index scales as ${\delta n_s}\simeq 0.4{\delta H_0\over
H_0}$ \cite{Ye:2021nej}, which suggests a scale-invariant
Harrison-Zeldovich spectrum ($n_s= 1$) for $H_0\sim 73$km/s/Mpc
\footnote{see \cite{Ye:2022afu} for Planck+BICEP/Keck dataset, in
which the $r-n_s$ contour for EDE was first showed ($r$ is the
tensor-to-scalar ratio).}. In
Refs.\cite{Jiang:2022uyg,Smith:2022hwi}, with
Planck+ACT+SPT+BAO+Pantheon dataset, similar results have also
been found. Here, we observed that the \textit{preference} for
$n_s=1$ is not altered by the inclusion of large-scale structure
DES-Y1 dataset (see recent \cite{Simon:2022adh} for BOSS dataset),
see Table-I,II. In this sense, the Hubble tension seems to hint
that we might live in a scale-free Universe, so it is interesting
to think about how $n_s=1$ would dramatically impact our
understanding on the primordial Universe and inflation
\cite{Ye:2021nej,DAmico:2021fhz,Kallosh:2022ggf,Ye:2022efx,Lin:2022gbl,Ageeva:2022asq,DAmico:2022agc}.

\section*{Acknowledgments}

We thank Gen Ye, Jun-Qian Jiang for helpful discussions. HW is
supported by UCAS Undergraduate Innovative Practice Project. YSP
is supported by the NSFC, No.12075246 and by the Fundamental
Research Funds for the Central Universities. We acknowledge the
use of publicly available codes AxiCLASS
(\url{https://github.com/PoulinV/AxiCLASS}) and classmultiscf
(\url{https://github.com/genye00/class_multiscf.git}).

\end{document}